\documentclass[prb,aps,twocolumn,amsmath,amssymb,floatfix,showpacs,
superscriptaddress]{revtex4}

\usepackage[dvips]{graphics}
\usepackage{color}
\definecolor{dred}{rgb}{0,0,0.6}

\begin{document}

\title{\textcolor{dred}{Circulating persistent current and induced magnetic 
field in a fractal network}}

\author{Srilekha Saha Saha}

\affiliation{Condensed Matter Physics Division, Saha Institute of Nuclear
Physics, Sector-I, Block-AF, Bidhannagar, Kolkata-700 064, India}

\author{Santanu K. Maiti}

\email{santanu.maiti@isical.ac.in}

\affiliation{Physics and Applied Mathematics Unit, Indian Statistical
Institute, 203 Barrackpore Trunk Road, Kolkata-700 108, India}

\author{S. N. Karmakar}

\affiliation{Condensed Matter Physics Division, Saha Institute of Nuclear
Physics, Sector-I, Block-AF, Bidhannagar, Kolkata-700 064, India}

\begin{abstract}

We present the overall conductance as well as the circulating currents 
in individual loops of a Sierpinski gasket (SPG) as we apply bias voltage 
via the side attached electrodes. SPG being a self-similar structure, its 
manifestation on loop currents and magnetic fields are examined in various 
generations of this fractal and it has been observed that for a given 
configuration of the electrodes, the physical quantities exhibit certain 
regularity as we go from one generation to another. Also a notable feature 
is the introduction of anisotropy in hopping causes an increase in magnitude 
of overall transport current. These features are a subject of interest 
in this article.

\end{abstract}


\maketitle 

\section{Introduction}

Fractals aroused a lot of interest in the minds of physicists and 
mathematecians for several decades because of their amazing physical 
and geometrical properties. Diverse types of fractals are present in 
nature, and can also be generated by means of some recursive rules. 
Among them some are of deterministic type, lying between systems of 
perfect periodic order and completely random ones, making it easier 
to explain the basic features of fractals~\cite{fick}. Sierpinski Gasket 
(SPG) is one such deterministic self-similar fractal where every scaled 
version exactly resembles to the original one. With advancement of 
lithographic techniques experimental designing of such geometry has become 
tailor made~\cite{new}. Electron transport through SPG has become a topic 
of current research interest~\cite{fr1,fr2,fr3,Maiti1}. Articles on 
quantum transport in mesoscopic systems mostly discuss the overall conduction 
properties~\cite{orella1,tagami,ventra2,ventra1,aviram,nitzan1,skm1,woi}, 
while little interests have been paid to distribution of currents 
at the various junctions and branches of a quantum network. Some articles 
brought into light, current distributions within the quantum networks by 
addressing possible interference effects arising from the multiple 
conduction pathways~\cite{nishi,suka}. While examining current 
distributions in systems attached to electrodes, possibility of such 
currents giving rise to circular currents have been reported~\cite{cir1}. 
But a proper definition of circular current came a little later~\cite{cir3}. 
Like the persistent current in isolated mesoscopic ring as predicted by 
B\"{u}ttiker~\cite{Butti} and his co-workers in $1983$, this circular 
current also persists as long as bias voltage is maintained. The behavior 
of persistent current in different geometries including fractals was 
usually explored only in presence of flux~\cite{gefen,Ore1,san1,san2,ambe,
schm1,schm2}, 
though the detail phenomena in individual loops of a quantum network need 
to be explored. Idea of persistent current arising as a direct consequence 
of Aharonov-Bohm (AB) effect led mesoscopic physics to the forefront of 
condensed matter physics. This self-sustaining current which originates 
in absence of external bias is a purely mesoscopic phenomenon and gets 
suppressed with increase in system size beyond the phase coherence length. 
\begin{figure}[ht]
{\centering \resizebox*{3.5cm}{2.2cm}{\includegraphics{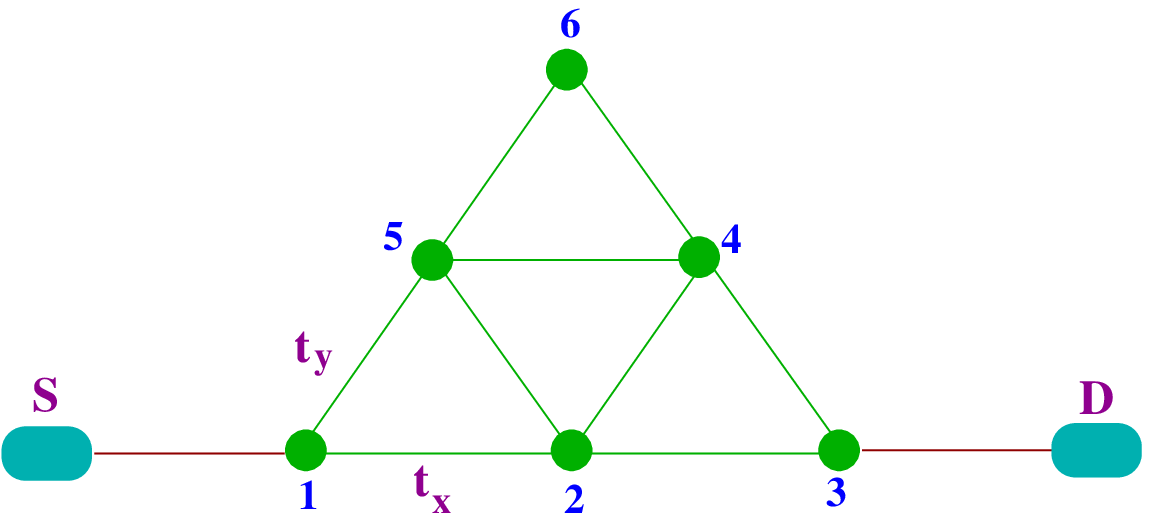}} 
\vskip 0.2cm 
\resizebox*{5cm}{3.2cm}{\includegraphics{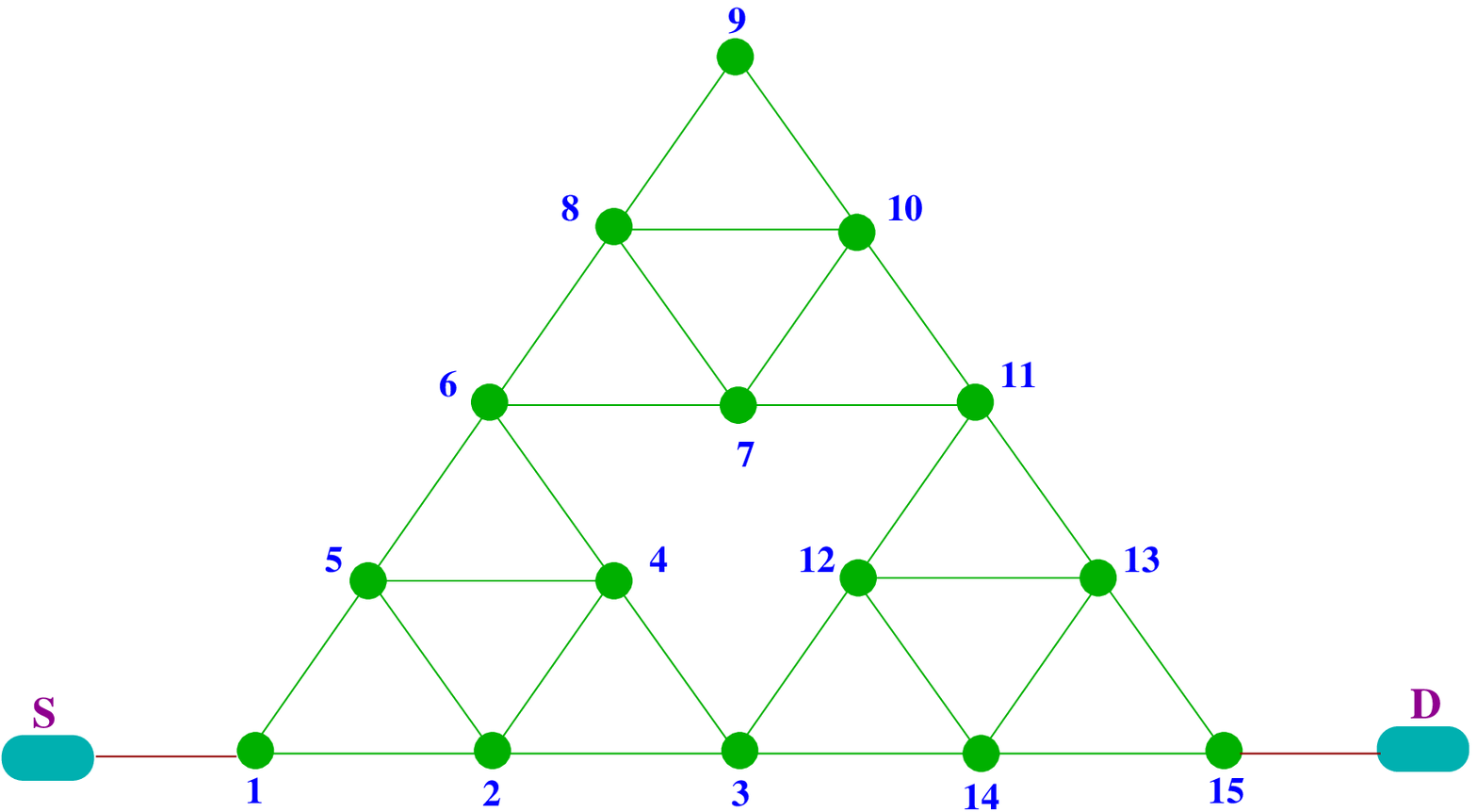}} 
\vskip 0.2cm
~\resizebox*{7.5cm}{6.2cm}{\includegraphics{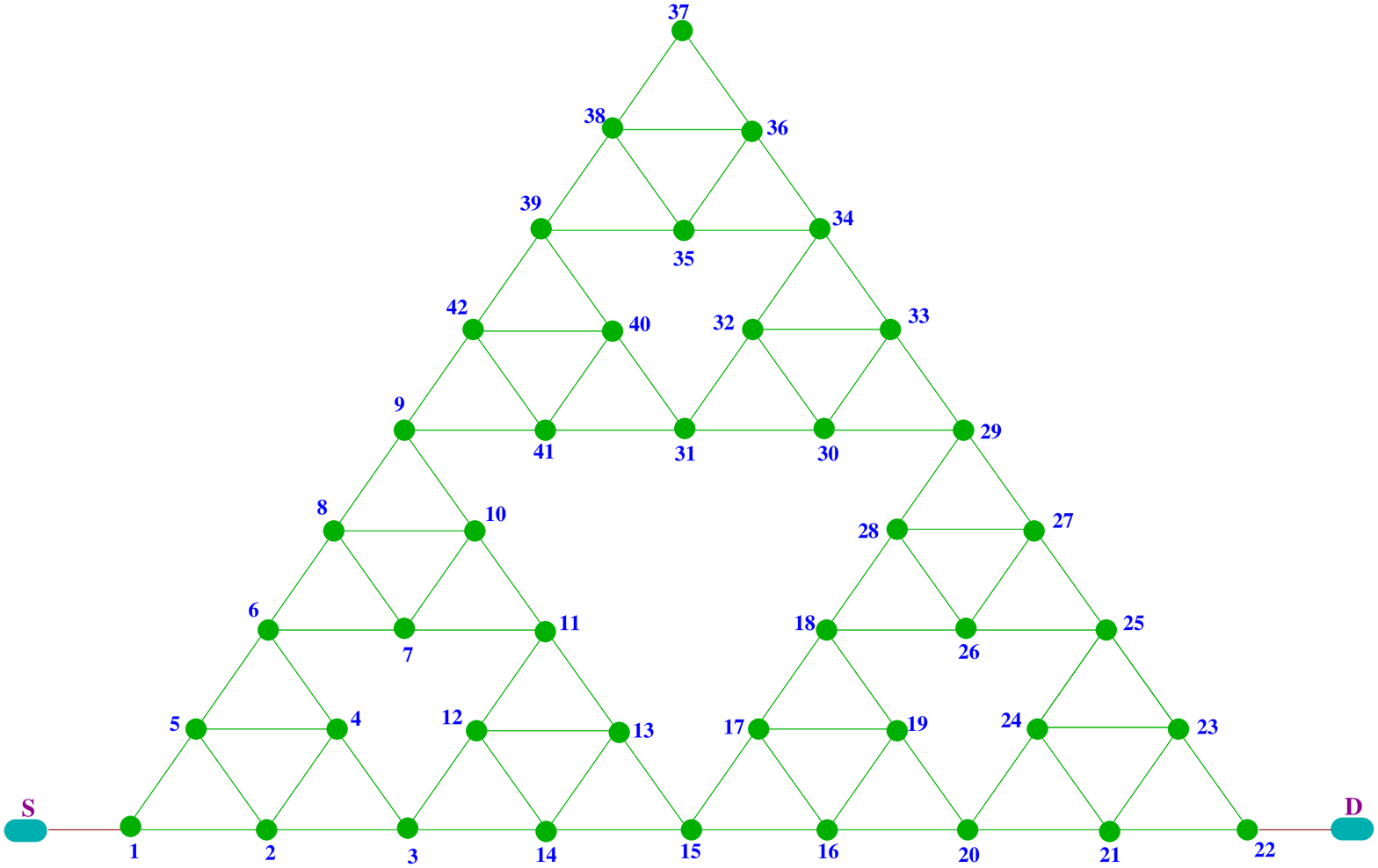}}\par}
\caption{(Color online). $1$st, $2$nd and $3$rd generations (top to bottom)
of a SPG fractal. Different numbers (viz, $1$, $2$, $3$, $\dots$) represent 
lattice sites of the network.}
\label{spg1}
\end{figure}
Some recent theoretical works have pointed out the fact that persistent 
current can be generated by circularly polarized light, twisted light and 
shaped photon pulses~\cite{l1,l2,l3,l4,l5,l6}. 
This equilibrium current arises not only in isolated geometries but also 
appears in loops connected to external electrodes. Persistent current was 
observed experimentally in closed systems~\cite{levy,mailly,chand}, but 
experimental demonstration of its counterpart in open system has not been 
observed till now to the best of our knowledge. Persistent current in 
presence of bias voltage can be observed in loop structures only, where
a loop is connected to external reservoirs.
Another notable feature that came into limelight is the fact that this 
circular current in turn gives rise to a magnetic field at the center of 
the loop and this field can be utilized to design quantum 
devices~\cite{cir3}. Orientation of a spin placed at the center of a loop 
geometry can be controlled by the magnetic field produced by circular 
current and this idea can be applied to generate spin based quantum 
devices. The following works dealt with some ways of controlling magnetic 
field externally. For example, Cho {\em et al.}~\cite{Cho} studied the 
behavior of coherent electron transport in a double quantum dot system 
and found that it acted as a magnetically polarized device due to quantum 
interference of electrons within the closed path of the device. They also 
showed that by varying the energy level of each dot, magnetic states of 
the device can be made up, down or non-polarized. In another work~\cite{Lidar} 
it has been depicted how an infinite array of parallel current carrying 
wires can produce exponentially localized magnetic field with peak value 
reaching upto $10\;$mT. A major drawback of this set-up is the issue 
of heating and 
loosing of coherence, and the device has been found to work well below 
the temperature $\ll 2.4\;$mK. In report by Pershin {\em et al.}~\cite{Per}, 
they showed how phase locked infra-red laser pulses can be used to generate 
local magnetic field at the ring center. Magnetic field of around $3\;$mT 
was induced at the center and it is quite comparable to that obtained in the 
preceding reference. In $2012$ Anda {\em et al.}~\cite{Anda} have provided 
the evidence of circular current in a ring with two quantum dots embedded 
in it and coupled to two electrodes, but nothing has been mentioned regarding 
the magnetic field induced by this current. In other recent work~\cite{san3} 
it has been shown how the local magnetic field can be regulated in a 
conducting mesoscopic ring subjected to an in-plane electric field. Though
few propositions have been made to scrutinize current distributions 
including circular current and associated magnetic field in different 
bridge systems, but most of these works are associated with simple loop
geometries and no one has addressed these characteristics in any simple
and/or complex network geometries. SPG fractal can be a suitable example of
it which may exhibit interesting patterns by virtue of the self-similarity. 
The main fact presented in this paper is variation of currents and induced 
magnetic field in each triangle of a SPG network with applied voltage, and 
how the nature of 
circular current varies with generations. An interesting feature obtained 
from the model is the mirror symmetry of currents in each loop on either 
side of an imaginary vertical line drawn perpendicular to the line connecting
the leads for the positions as given in Fig.~\ref{spg1} for any generation.
We also present how the change in position of such leads affect circular 
currents and induced magnetic field. Our analysis can be useful to extract
important information in other related self-similar structures.

The paper is arranged as follows. Section II gives the theoretical 
formulation with the numerical results discussed in subsequent Section III, 
which include the analysis of transmission spectra and transport current, 
hence follows the discussion of circular currents in individual loops and
the associated magnetic fields. Finally, we conclude in the last Section IV.

\section{Model and Theoretical Formulation}

We begin by referring to Fig.~\ref{spg1} depicting different generations of 
a SPG network connected to source ($S$) and drain ($D$) electrodes with the 
hopping in angular and horizontal directions denoted by $t_x$ and $t_y$,
respectively. The filled green circles represent atomic sites and we 
describe the entire system by using a tight-binding formalism. The 
Hamiltonian of the entire system reads,
\begin{equation}
\mathcal{H}=H_S+ H_D + H_{SPG} + H_{S,SPG}+ H_{D,SPG}
\label{eq1}
\end{equation}
First two terms correspond to the Hamiltonians for semi-infinite leads 
and can be expressed as
\begin{equation}
H_{S} + H_{D} = \sum \limits_{S,D} \left\{ \sum \limits_n \epsilon_0 
\alpha_n^\dag \alpha_n + 
\sum\limits_n t_0 [\alpha_{n+1}^\dag \alpha_n+h.c.] \right \}
\label{eq2}
\end{equation}
where $\epsilon_0$ is the on-site energy and $t_0$ represents 
nearest-neighbor hopping strength in electrodes. Creation and annihilation 
operators for an electron in the $n$th site are labeled by $\alpha_n^\dag$ 
and $\alpha_n$, respectively.

The Hamiltonian for the SPG triangle is
\begin{equation}
H_{SPG}= \sum \limits_i \epsilon_i c_i^\dag c_i + 
\sum \limits_{\langle i j \rangle} t_{ij} \left[c_i^\dag c_j + h.c. \right]
\label{eq3}
\end{equation}
where $c_i^\dag$ and $c_i$ stand for the creation and annihilation operators, 
while $\epsilon_i$ and $t_{ij}$ are the on-site energy and nearest-neighbor 
hopping integral, respectively, for the SPG. We take $\epsilon_i=\epsilon$ 
and $t_{ij}=t_x$ or $t_y$ depending on whether electron is hopping in 
the horizontal or in angular direction as show in Fig.~\ref{spg1}.

Finally, the last two terms arise due to the coupling between semi-infinite 
leads and atomic sites of SPG network and explicitly we can write  
\begin{equation}
H_{S,SPG} + H_{D,SPG} = \tau_S [c_1^\dag \alpha_0 + h.c.] 
+ \tau_D [c_q^\dag \alpha_{n+1} + h.c.]
\label{eq4} 
\end{equation}
where, the source is coupled to site $1$ of the SPG via the hopping strength 
$\tau_S$ and the other lead is attached to site $q$, which is variable, 
through the hopping strength $\tau_D$.

\subsection{Evaluation of transport current}

In order to evaluate transport current due to flow of electrons from source 
to drain, we need to calculate two-terminal transmission probability 
$T(E)$. Using Green's function formalism $T(E)$ can be obtained assuming 
the transport to be taking place in the coherent regime. Thus we can express 
the single particle Green's function operator in the form defined by
\begin{equation}
G=(E - H + i\eta)^{-1}
\label{eq5} 
\end{equation}
Introducing the contact self energies $\Sigma_S$  and  $\Sigma_D$
which incorporate the effect of couplings between (SPG) and the 
semi-infinite leads, the problem of determining $G^r$ in full Hilbert 
space can be transformed to the reduced Hilbert space spanned by the 
system itself, and we can write the effective Green's function~\cite{datta} 
\begin{equation}
\mathcal{G}^r=(E- H_{SPG} - \Sigma_S - \Sigma_D)^{-1}
\label{eq6} 
\end{equation}
In terms of retarded and advanced Green's functions, the two-terminal 
transmission probability of an electron can be written as~\cite{datta}
\begin{equation}
T(E) = Tr[\Gamma_S \mathcal{G}^r \Gamma_D \mathcal{G}^a]
\label{eq7}
\end{equation}
where $\mathcal{G}^a= [\mathcal{G}^r]^\dag$, $\Gamma_S$ and $\Gamma_D$ 
represent the coupling matrices.

This transmission function can be utilized to find junction 
current $I_T$ as a function of bias voltage following the 
relation~\cite{datta}
\begin{equation}
I_T = \frac{2 e}{h} \int \limits_{-\infty}^\infty T(E) 
\left[f_S (E) - f_D (E)\right]\;dE
\label{eq11}
\end{equation}
where, $f_S (E)$ and $f_D (E)$ are the Fermi functions of the source and 
drain, respectively. At absolute zero temperature this relation simplifies 
to
\begin{equation}
I = \frac{2 e}{h} \int \limits_{E_F -eV/2}^{E_F + eV/2} T(E)\; dE
\label{eq12}
\end{equation}
where, $E_F$ is the equilibrium Fermi energy and we set it to zero throughout
the calculations.

\subsection{Circular current and associated magnetic field}

\begin{figure}[ht]
{\centering \resizebox*{4cm}{3.5cm}{\includegraphics{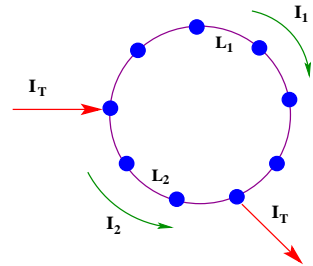}}\par}
\caption{(Color online). Circular loop connected to two leads exhibiting
currents in upper and lower arms of the ring.}
\label{ring}
\end{figure}

To determine circular current in each of the plaquettes of a SPG network, 
let us first analyze the current distribution in a circular loop as shown 
in Fig~\ref{ring}. 
The diagram shows that the current flowing in the electrodes being $I_T$, 
while $I_1$ and $I_2$ are currents passing through the upper and lower 
branches of the circular geometry. Net current can be expressed in terms of 
branch currents as
\begin{equation}
I_T = I_1 -I_2
\label{eq13} 
\end{equation}
Here we have assumed clockwise direction of current to be positive. The 
above expression (Eq.\ref{eq13}) can be rearranged in the form~\cite{cir3} 
$I_T = (I_1 - I_c)- (I_2 - I_c)= I_1^{tr} - I_2^{tr}$, where $I_c$ is the 
current circulating in the ring, and being persistent in nature, arises 
solely due to the bias voltage.

This self-sustaining current can be expressed as
\begin{equation} 
I_c=\frac{I_1 L_1 + I_2 L_2}{L}
\label{eq14}
\end{equation}
where, $L_1$ and $L_2$ are the lengths of the upper and lower arms and 
$L=L_1 + L_2$. This current does not contribute to the net current, only 
keeps on circulating in the loop. The formula can be generalized in case 
of more than two branches as
\begin{equation} 
I_c= \frac{\sum \limits_i {I_i L_i}}{L}
\label{eq15}
\end{equation} 
From the expressions of circular current it becomes clear that we need 
to evaluate the branch currents, which can be done using the Green's 
function formalism. First we calculate the bond current density~\cite{tag}
\begin{equation} 
J_{ij} = \frac{4e}{h} \mbox{Im}[(H_{SPG})_{ij} \mathcal{G}_{ij}^n]
\label{eq16}
\end{equation} 
We define $\mathcal{G}^n$ as the correlated Green's function given by 
the formula 
\begin{equation} 
\mathcal{G}^n = \mathcal{G}^r \Gamma_S \mathcal{G}^a
\label{eq17}
\end{equation}  
where, each term has the same meaning as mentioned earlier. Now, the bond 
current between nearest-neighbor sites $i$ and $j$ at absolute zero 
temperature is given by 
\begin{equation} 
I_{ij} = \int \limits_{E_F -eV/2}^{E_F + eV/2} J_{ij}(E)\; dE
\label{eq18}
\end{equation} 
Substituting the above expression in Eq.~\ref{eq14} or Eq.~\ref{eq15} 
depending upon the 
system, we finally evaluate circular currents in each triangle of SPG 
network. This circulating current gives rise to local magnetic field at 
the center of each plaquette and it can be analyzed using Biot-Savart's 
law~\cite{cir3}
\begin{equation} 
B = \sum \limits_{ij} \frac{\mu_0}{4 \pi} \int I_{ij} \frac{d\vec{r} \times 
(\vec{r}-\vec{r}^{\prime})}{|\vec{r}-\vec{r}^{\prime}|^3}
\label{eq19}
\end{equation} 
where, $\mu_0$ signifies the magnetic constant and $\vec{r^{\prime}}$ 
describes the position vector of the bond current element.

\section{Numerical Results and Discussion}

Based on above theoretical formulation we present below the results 
of our numerical calculations.

\subsection{Transmission spectra and transport current} 

\begin{figure}[ht]
{\centering \resizebox*{8.5cm}{11cm}{\includegraphics{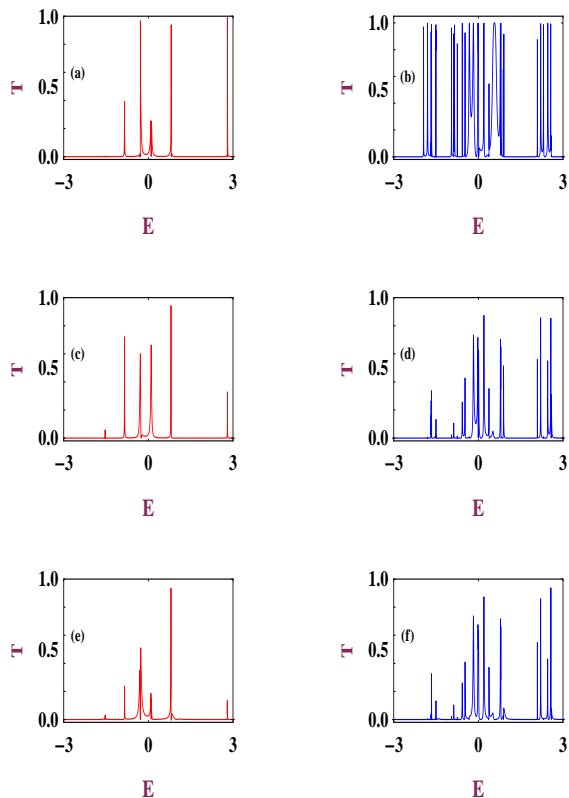}}\par}
\caption{(Color online). $T$-$E$ characteristics for a $3$rd generation 
SPG ($N=42$), where the left and right columns correspond to 
$t_x=t_y=1\;$eV and $t_x=1.0\;$eV, $t_y=0.7\;$eV, respectively. Here the
source is coupled to site $1$, while the drain is attached to three 
different sites $22$, $29$ and $9$ those are presented in the $1$st, 
$2$nd and $3$rd rows, respectively.}
\label{spg2}
\end{figure}

Figure~\ref{spg2} shows the two-terminal transmission probability $T$ as 
function of energy $E$ for a $3$rd generation SPG network. The three 
different rows correspond to three different positions of 
drain ($22$, $29$ and $9$) with source being kept 
fixed at position $1$. The left column corresponds to the case of isotropic 
hopping integral $t_x=t_y$, while the other column represents the anisotropic 
($t_x \ne t_y$) counterpart. It is evident from the spectra that anisotropy 
introduces more peaks in the transmission characteristics, thereby making 
the system more conducting in nature than the isotropic case. It is also 
observed from the $T$-$E$ characteristics, that the conducting nature 
is highest for the first row (which is even more prominent in 
Fig.~\ref{spg2}(b)) when leads are connected at two sides of the SPG, 
compared to other rows when drain lead is connected above 
the base line.

\begin{figure}[ht]
{\centering \resizebox*{8cm}{11cm}{\includegraphics{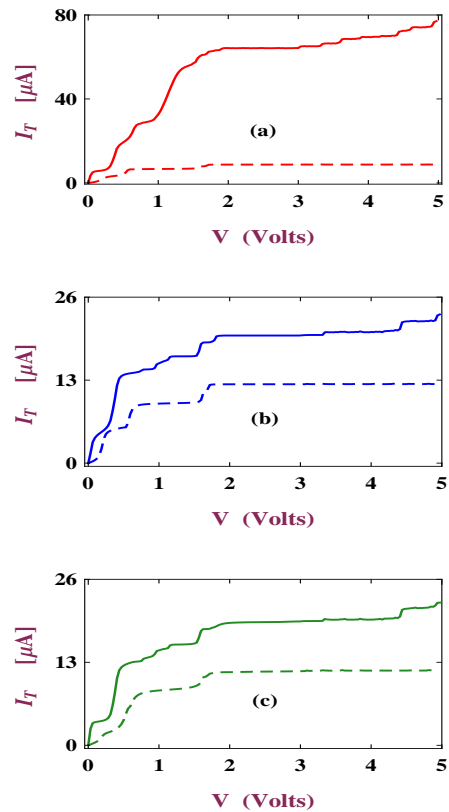}}\par}
\caption{(Color online). $I_T$-$V$ characteristics of a $42$-site SPG 
network where the dotted and solid lines correspond to isotropic hopping
($t_x=t_y=1\;$eV) and its anisotropic counterpart ($t_x=1\;$eV, 
$t_y=0.7\;$eV). Here the source is coupled to site $1$, 
while the drain is attached to three different sites 
those are: (a) $22$, (b) $29$ and (c) $9$.}
\label{spg3}
\end{figure}

From this result (Fig.~\ref{spg2}) we can make a sense about the nature 
of transport current and how anisotropy may result in an increase in 
magnitude of the current. The $I_T$-$V$ characteristics of the identical
SPG fractal are shown in Fig.~\ref{spg3} for three different positions of 
the drain ($22$, $29$, and $9$), as taken in Fig.~\ref{spg2}, 
connecting the source lead to the site $1$. The solid and dashed curves 
represent anisotropic and isotropic fractals respectively. These spectra 
clearly depict the increment of current for the anisotropic hopping case
following the transmission characteristics as presented in Fig.~\ref{spg2}. 
Physically anisotropy reduces
\begin{figure}[ht]
{\centering \resizebox*{8.5cm}{11cm}{\includegraphics{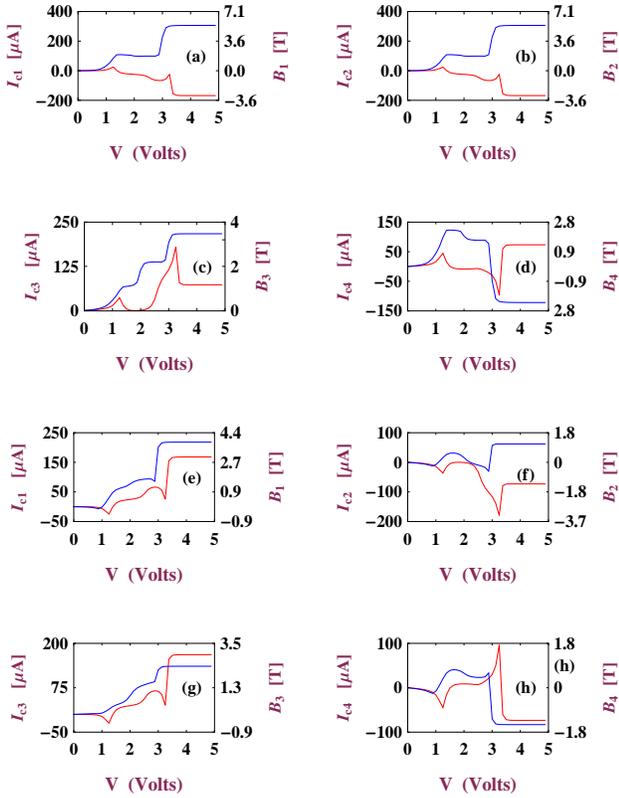}}\par}
\caption{(Color online). Variation of circular current $I_{cn}$ ($n$ 
corresponds to individual triangles) and associated magnetic field $B_n$ 
with bias voltage for each plaquettes of 1st generation SPG network. Here 
the red curves are for $t_x=t_y=1\;$eV while blue lines are for $t_x=1\;$eV, 
$t_y=0.7\;$eV. For (a)-(d) leads are connected at $1$ and $3$ sites, while
for (e)-(h) they are coupled to sites $1$ and $6$.}
\label{spg4}
\end{figure} 
the possibility of destructive quantum interference in the various loops 
of the SPG network and hence increases the overall transport current. We 
also find that the junction current becomes maximum when leads are attached 
\begin{figure}[ht]
{\centering \resizebox*{8cm}{4cm}{\includegraphics{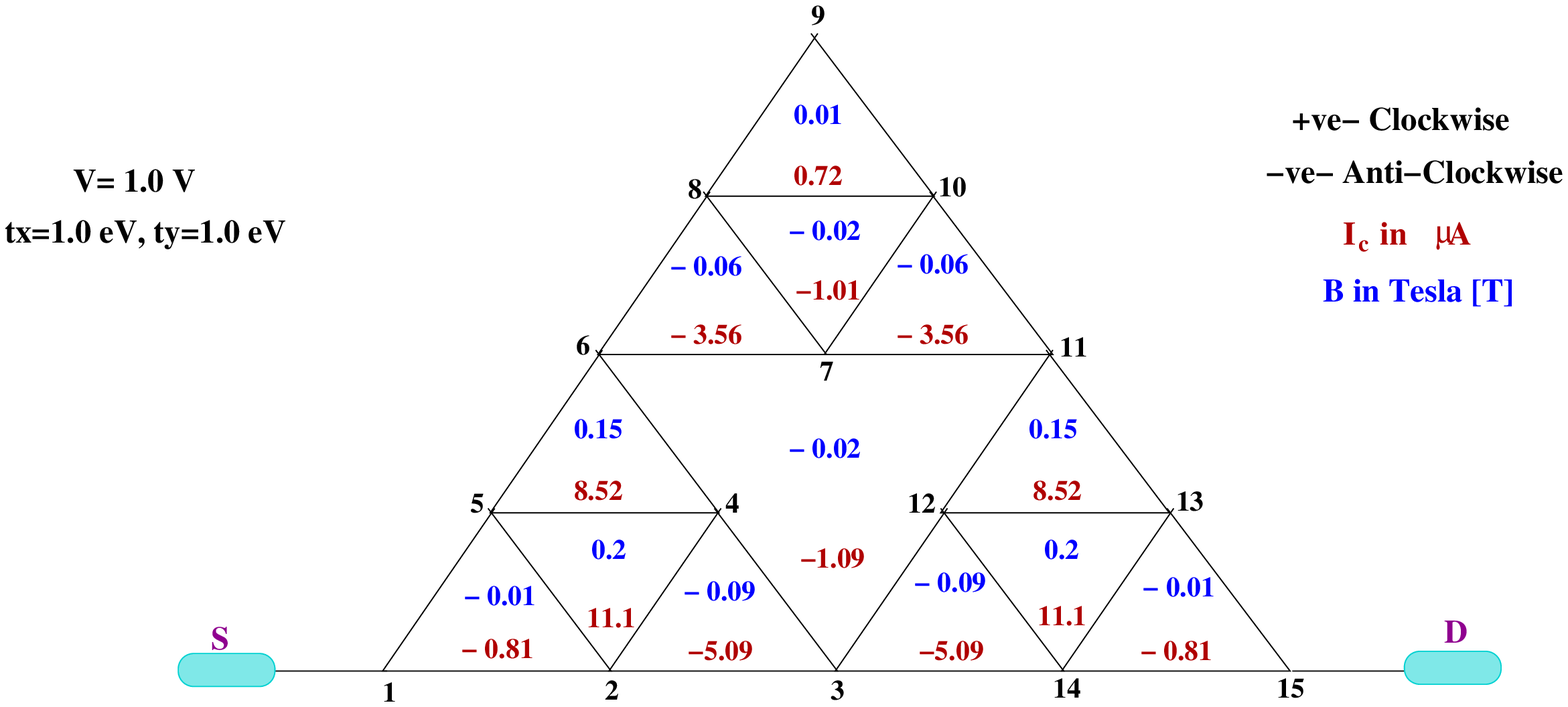}}
\vskip 0.3cm
\resizebox*{8cm}{4cm}{\includegraphics{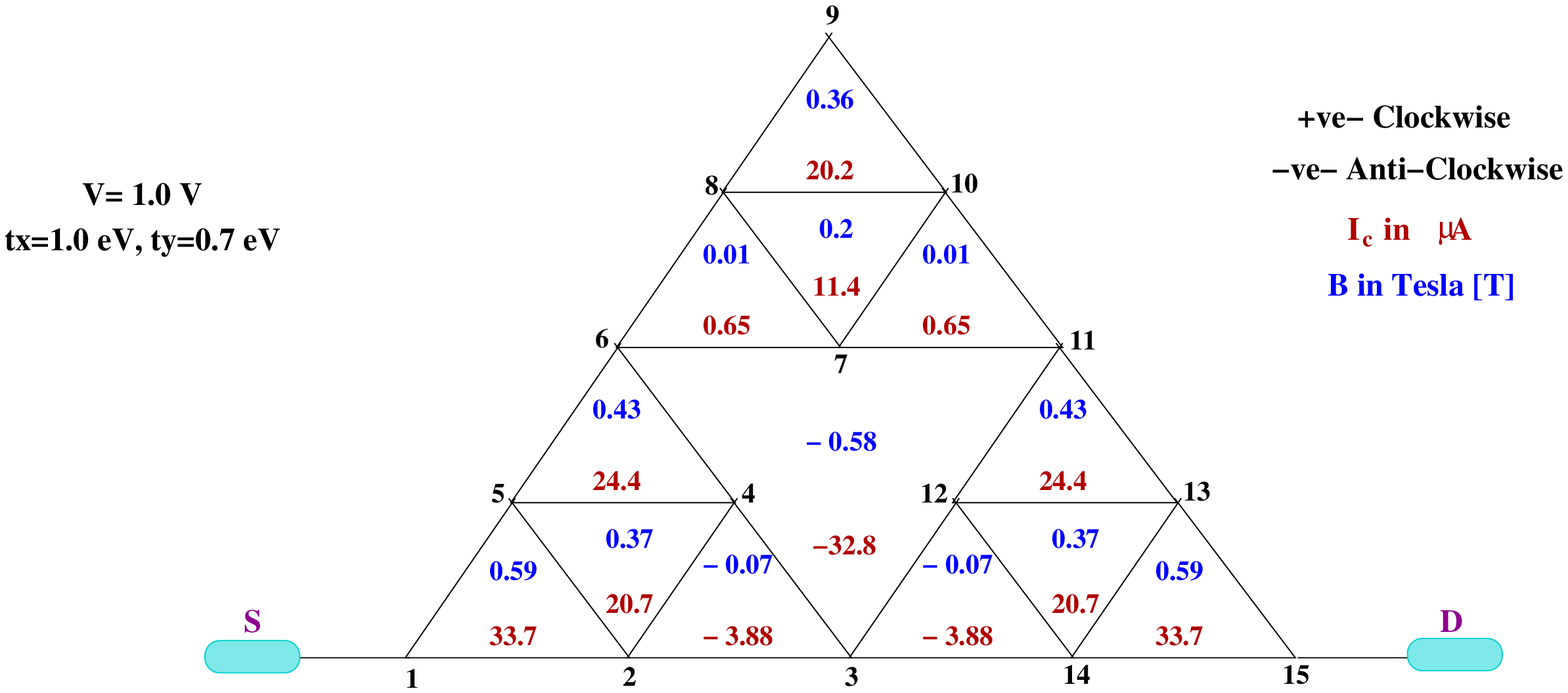}}\par}
\caption{(Color online). Values of circular currents and associated 
magnetic fields at individual plaquettes of a $2$nd generation SPG ($N=15$)
network. All the other physical parameters are clearly mentioned in the
figure.}
\label{spg5}
\end{figure} 
\begin{figure}[ht]
{\centering \resizebox*{8cm}{4cm}{\includegraphics{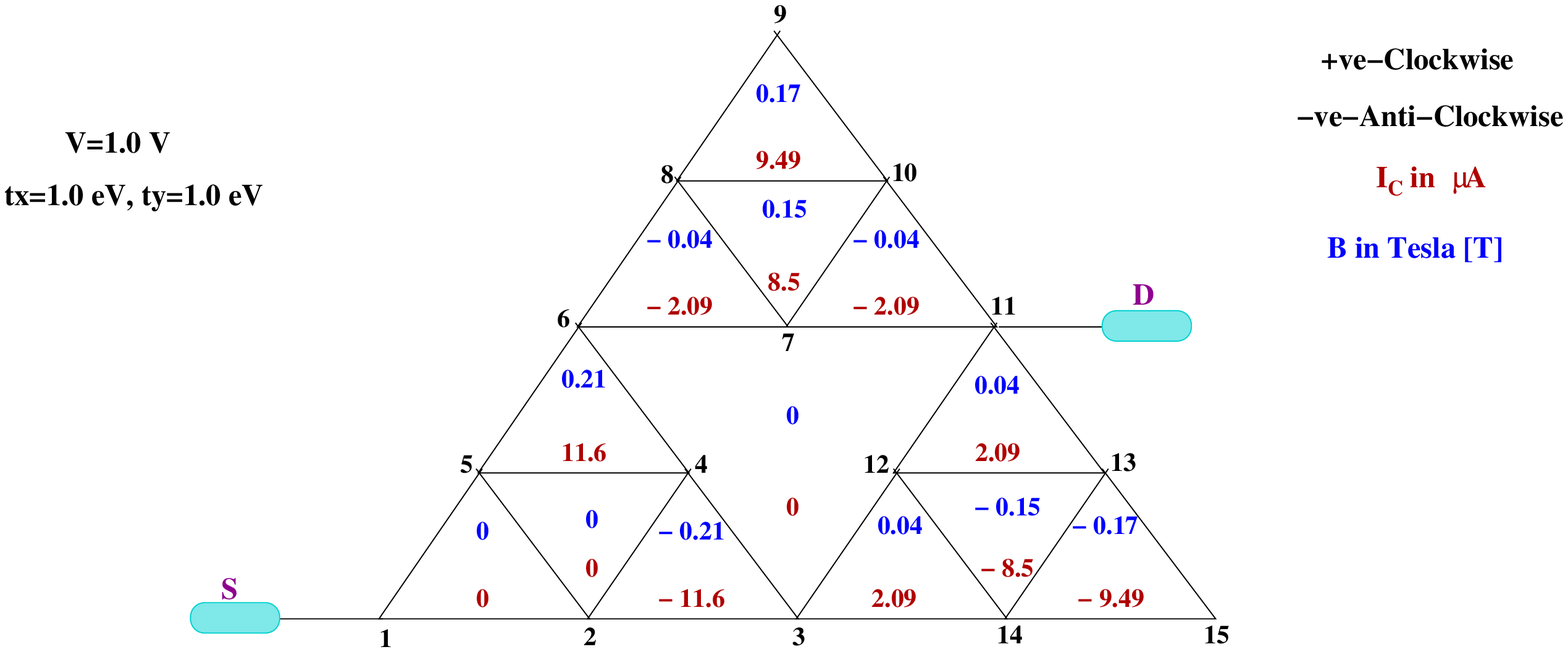}}
\vskip 0.3cm
~~~\resizebox*{8cm}{4cm}{\includegraphics{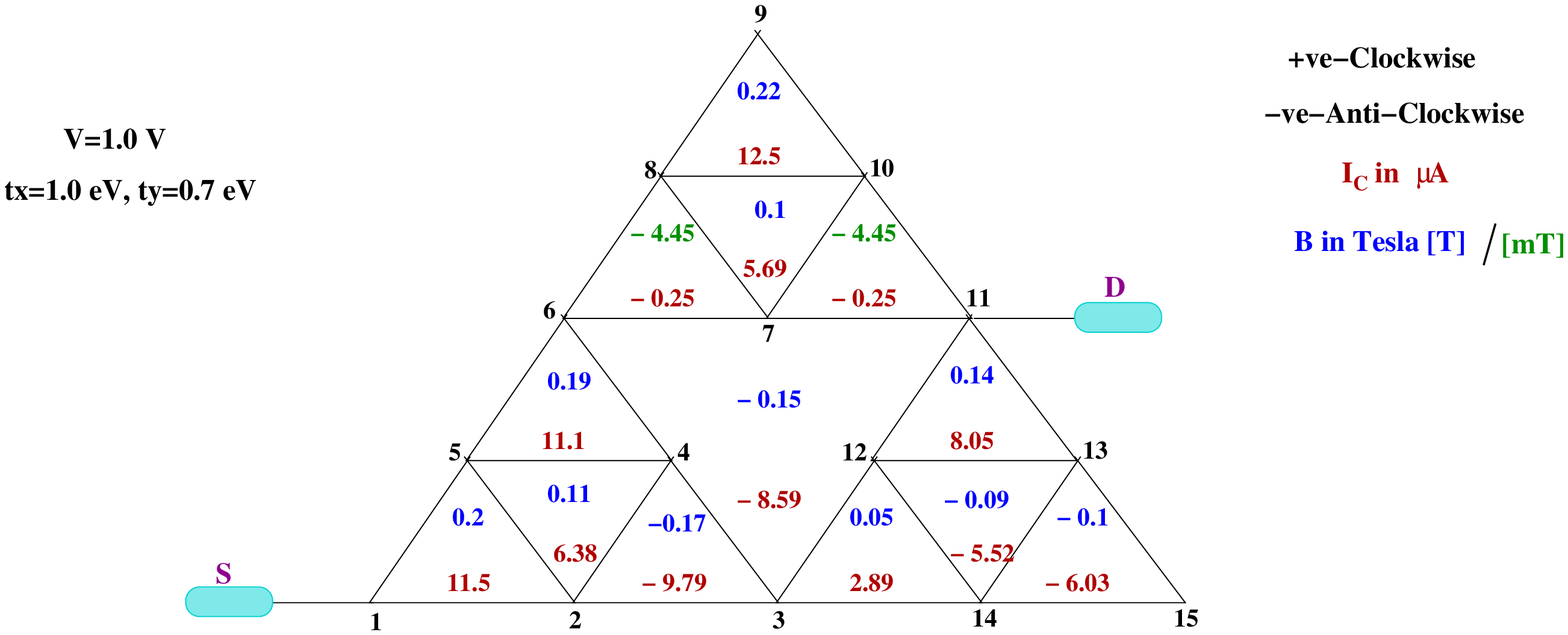}} \par}
\caption{(Color online). Same as Fig.~\ref{spg5} with different 
source-SPG-drain configuration.}
\label{spg6}
\end{figure}
on two sides of the base of SPG. The reason behind it is that the chances 
of destructive interference is less for this SPG-lead configuration 
($1$-$22$) compared to the other configurations, viz, $1$-$29$ and $1$-$9$.

\subsection{Circular currents and magnetic field} 

So far we have discussed transport currents through SPG, and, now we try 
to analyze the behavior of currents in each triangular loop of this
network.
\begin{figure}[ht]
{\centering \resizebox*{8cm}{5.5cm}{\includegraphics{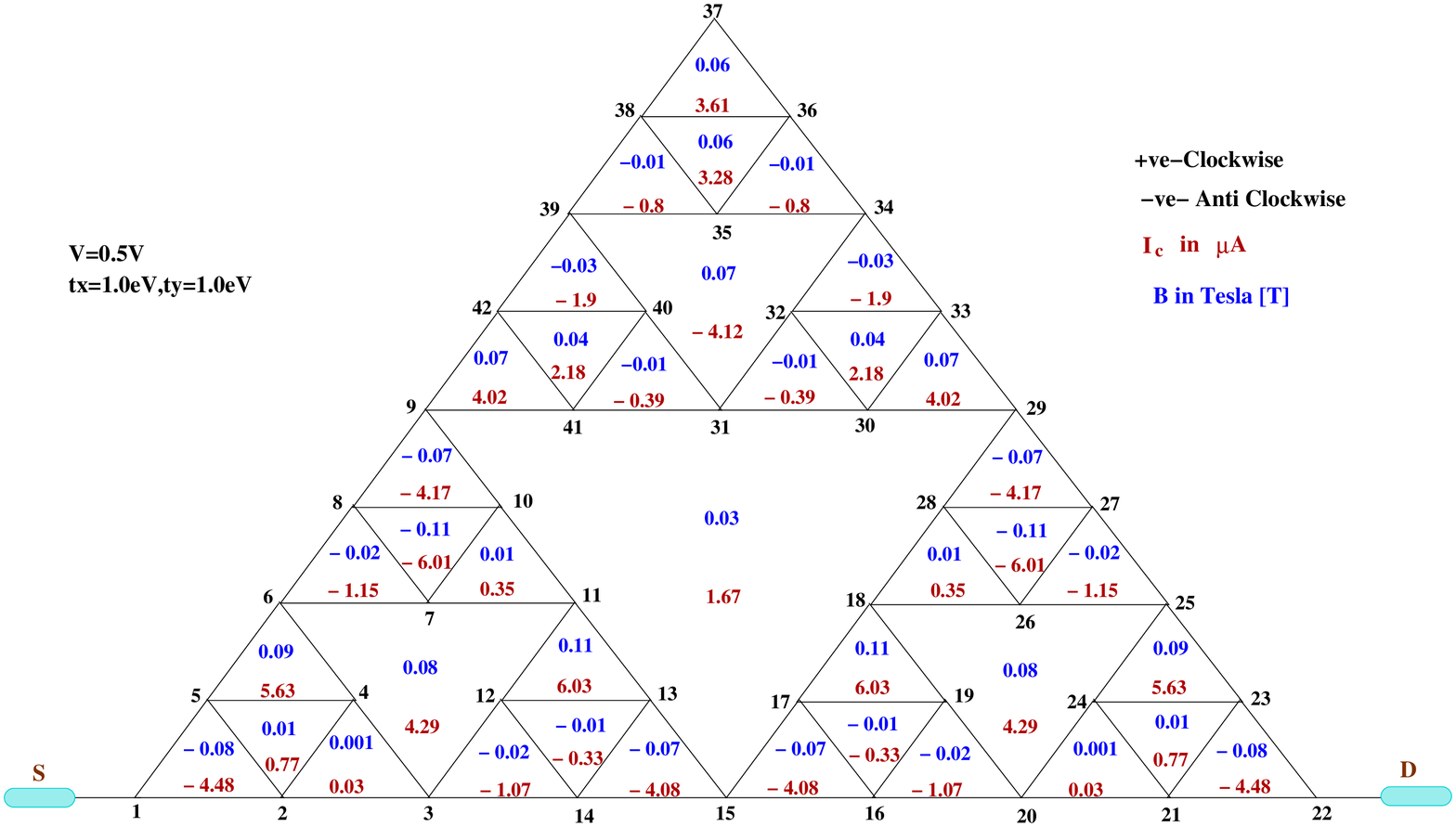}}
\vskip 0.3cm
\resizebox*{8cm}{5.5cm}{\includegraphics{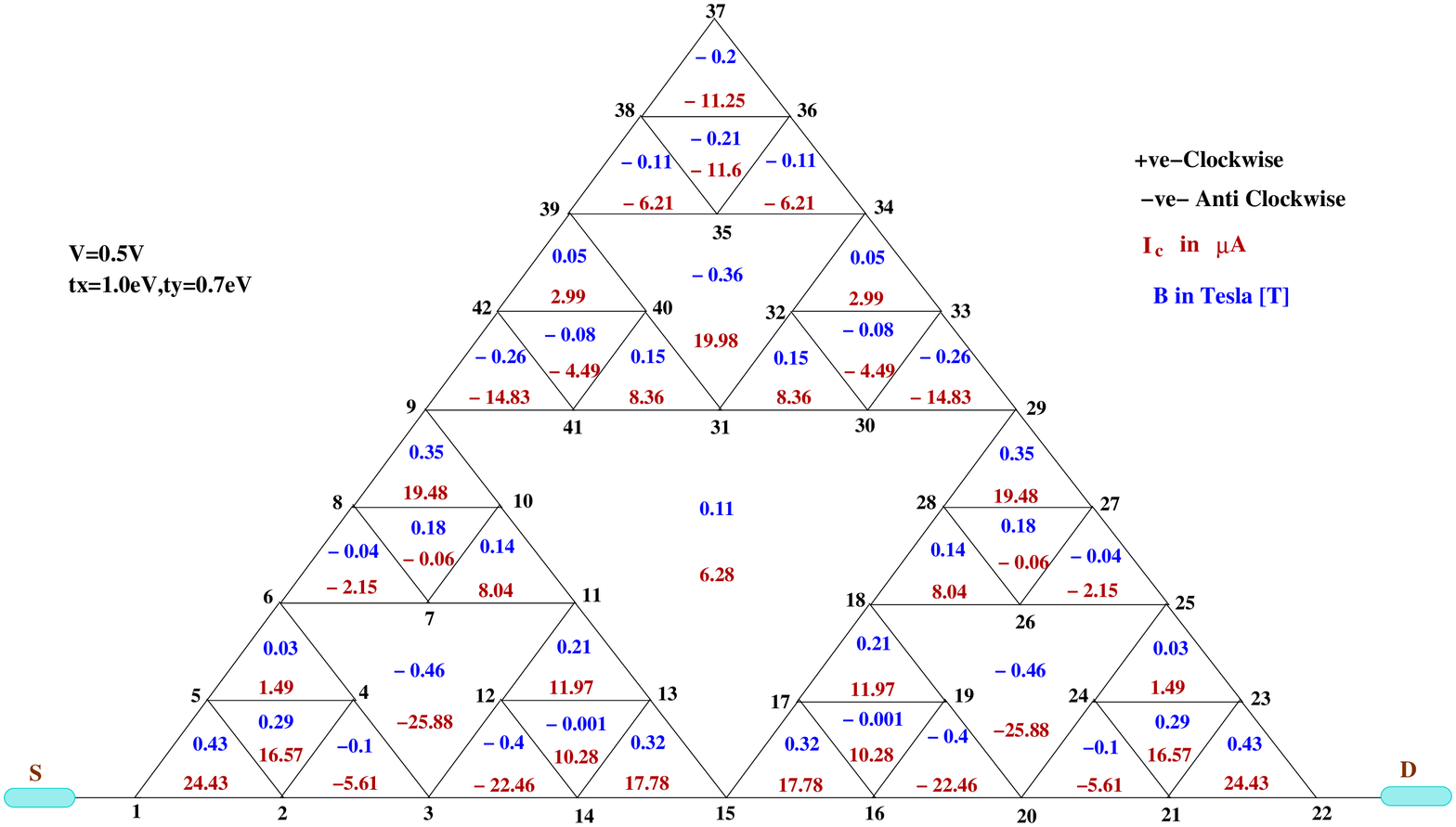}}\par}
\caption{(Color online). Values of circular currents and associated magnetic
fields at individual plaquettes of a $3$rd generation SPG ($N=42$) network.
All the other physical parameters are mentioned in the figure.}
\label{spg7}
\end{figure}
Figure~\ref{spg4} exhibits the variation of circular current as well as 
magnetic field originating at the center due to this current in each 
plaquette for a smallest SPG fractal ($1$st generation) where $N=6$. 
The red and blue curves correspond to the isotropic and anisotropic cases 
respectively. Variation of circular current and magnetic field with bias 
voltage in individual loops can be seen clearly. Both these quantities
behave in a similar manner, apart from a scale factor, and they cannot be 
predicted to be either increasing or decreasing consistently due to the
introduction of anisotropy unlike the case of transport current. The
spectra (a)-(d) indicate the situations where leads are connected at sites 
$1$ and $3$, while in (e)-(h) the results are shown when the leads are 
connected at the sites $1$ and $6$. Quite interestingly we find that the 
circular currents $I_{c1}$ (measured for the triangle connecting sites 
$1$, $2$ and $5$) and $I_{c2}$ (measured for the triangle connecting sites 
$2$, $3$ and $4$), and associated magnetic fields, $B_1$ and $B_2$, exhibit 
exactly identical variation with external bias voltage for both the 
isotropic and anisotropic cases yielding a mirror symmetry about an 
\begin{figure}[ht]
{\centering \resizebox*{8cm}{5.5cm}{\includegraphics{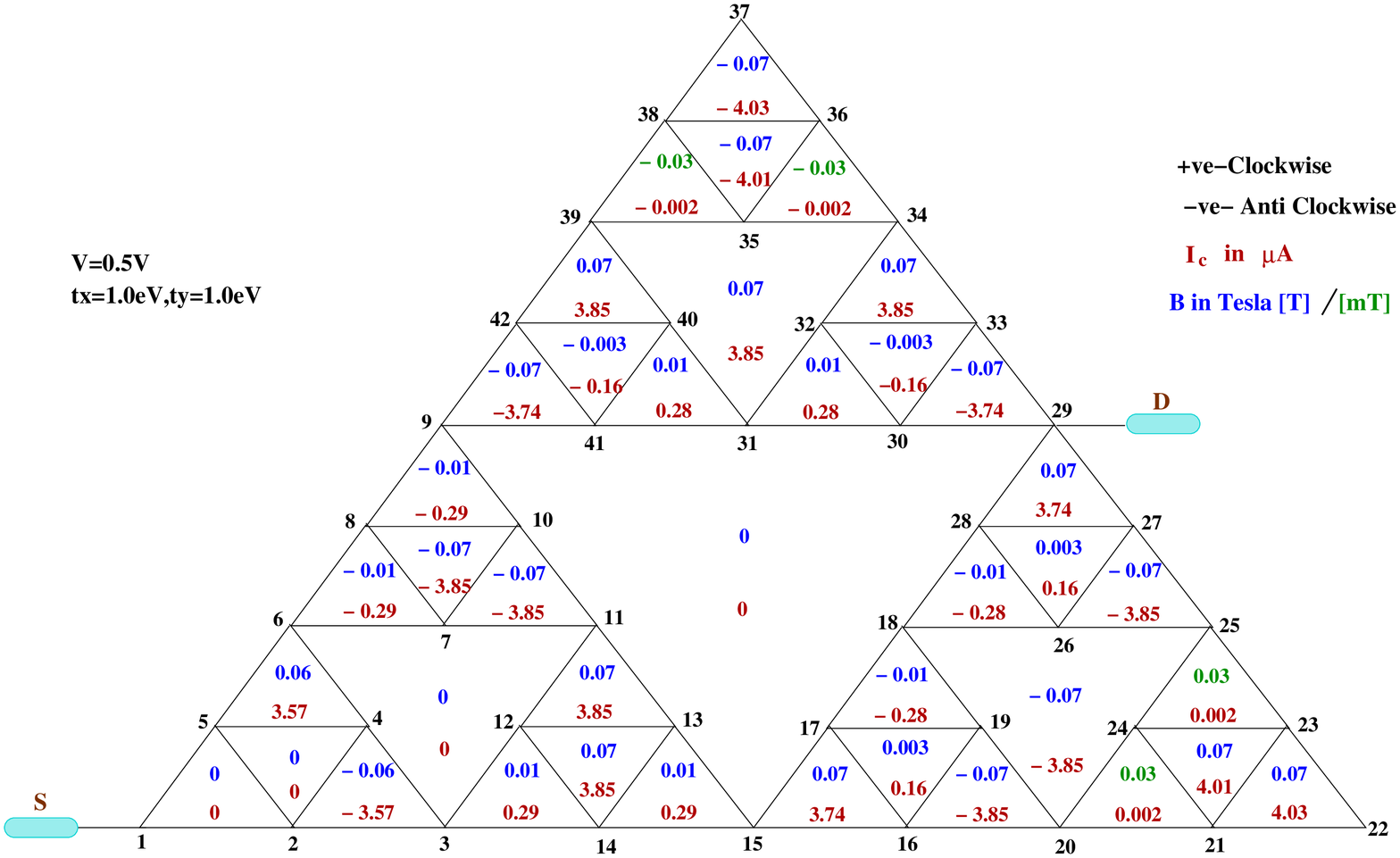}} 
\vskip 0.3cm
\resizebox*{8cm}{5.5cm}{\includegraphics{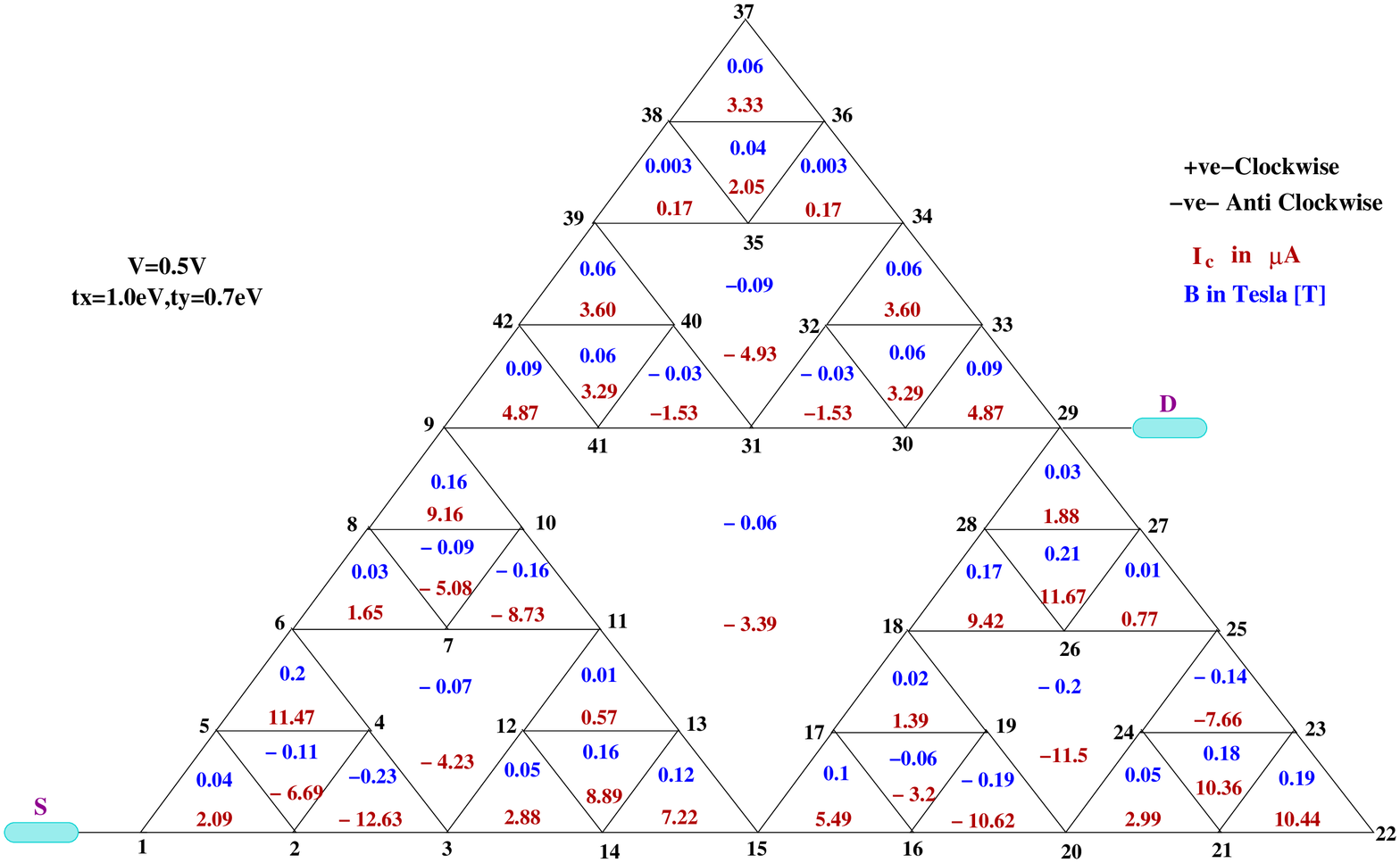}} \par}
\caption{(Color online). Same as Fig.~\ref{spg7} with different 
source-SPG-drain configuration.}
\label{spg8}
\end{figure} 
imaginary vertical axis dividing the SPG into two equal halves. This unique 
feature is observed only when leads are connected on either sides of the 
bases of SPG e.g., at positions $1$ and $3$ for this particular generation 
of the fractal. Analogous feature is also observed when leads are coupled 
to the sites ($1$, $15$) and ($1$, $22$) of $2$nd and $3$rd generations, 
respectively, and even valid for any higher generation. This mirror 
symmetry gets disturbed when the drain lead is connected 
elsewhere. 

In the rest of our discussion we point out some interesting facts regarding 
the magnitudes of current and magnetic field for $2$nd and $3$rd generation 
\begin{figure}[ht]
{\centering \resizebox*{7cm}{3cm}{\includegraphics{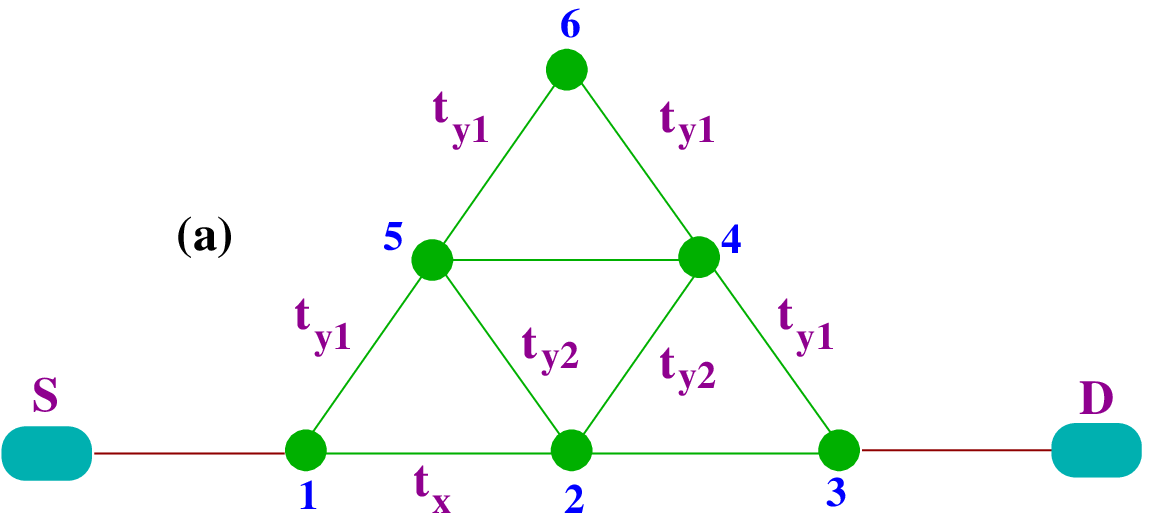}} 
\vskip 0.3cm
\resizebox*{8.5cm}{3cm}{\includegraphics{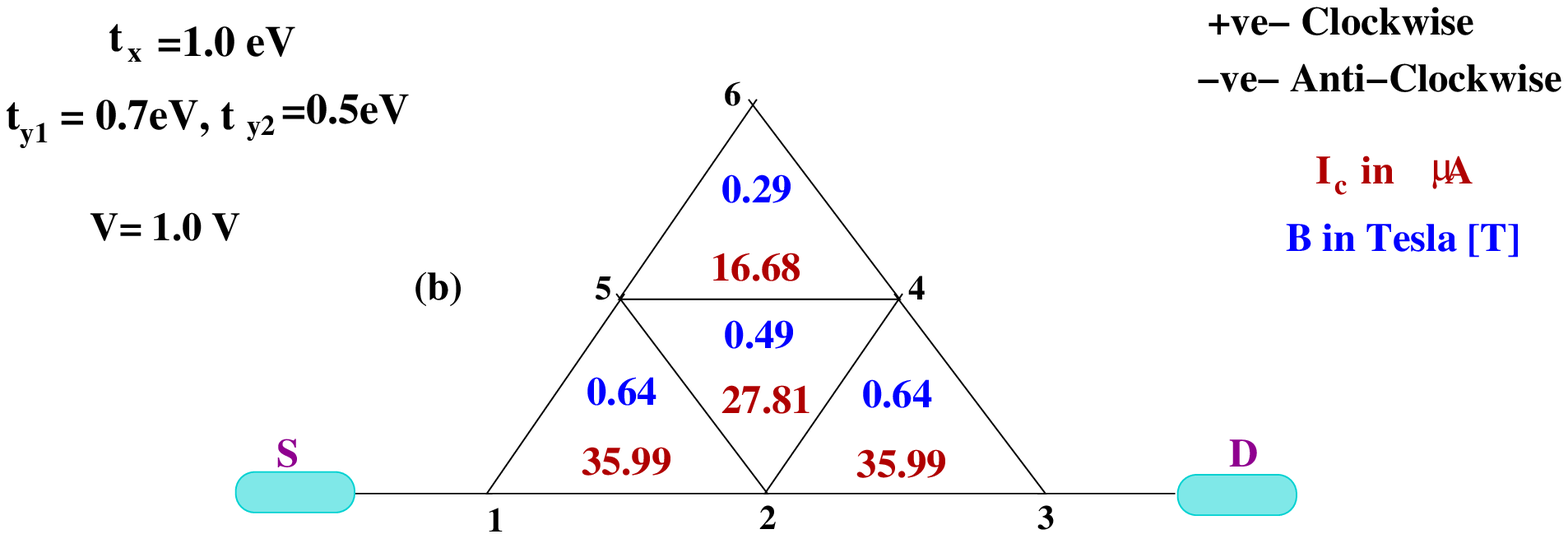}} \par}
\caption{(Color online). Values of circular currents and
associated magnetic fields in different triangular plaquettes for a
$6$-site SPG network subjected to three different hopping integrals
($t_x$, $t_{y1}$ and $t_{y2}$). The schematic diagram of the network is
shown in (a), while in (b) the numerical results are presented those
are computed for a particular set of parameter values as given in this
spectrum.}
\label{nfig1}
\end{figure} 
\begin{figure}[ht]
{\centering \resizebox*{7cm}{3cm}{\includegraphics{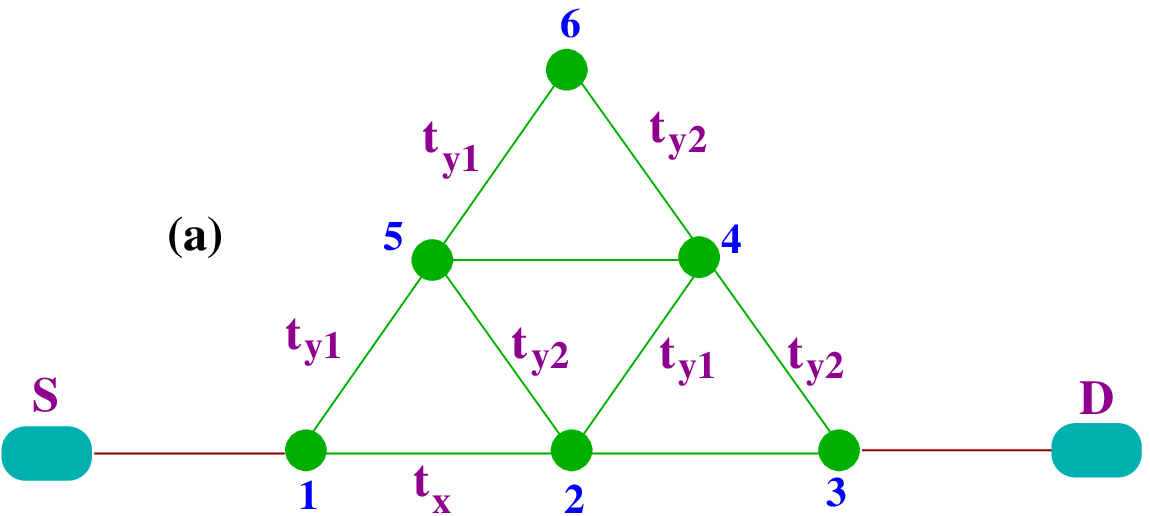}} 
\vskip 0.3cm
\resizebox*{8.5cm}{3cm}{\includegraphics{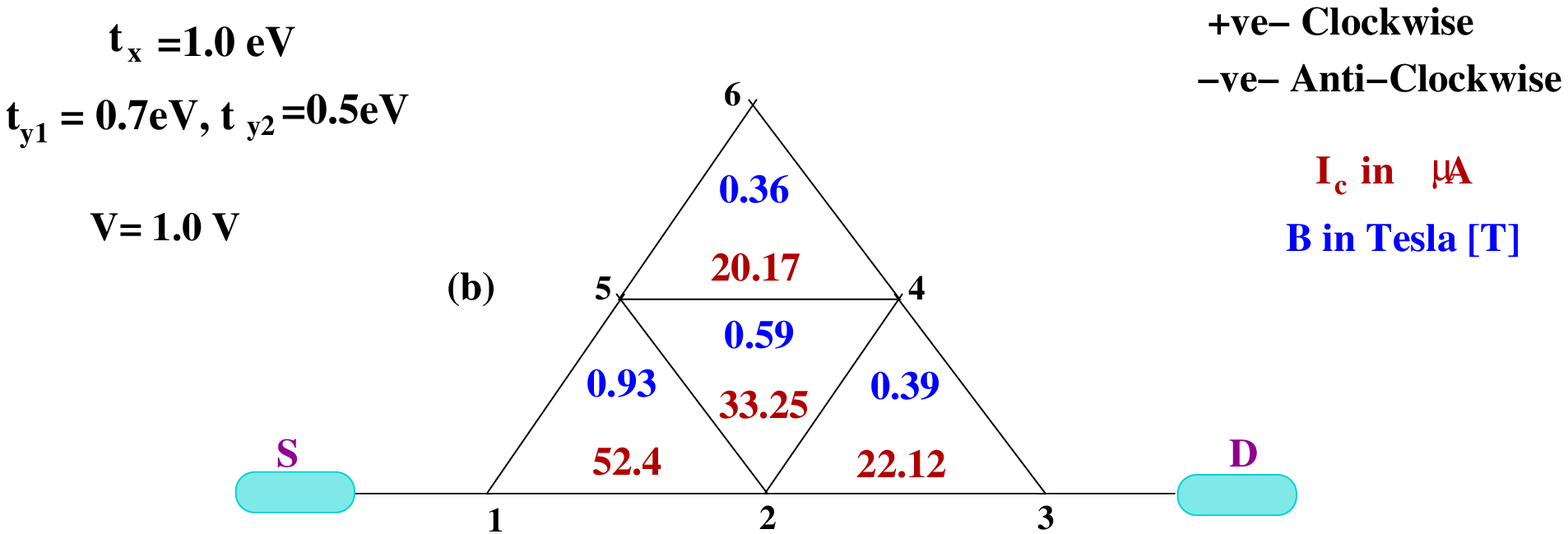}} \par}
\caption{(Color online). Same as Fig.~\ref{nfig1} with a different 
arrangement of two hopping integrals $t_{y1}$ and $t_{y2}$.}
\label{nfig2}
\end{figure} 
fractals. Figures~\ref{spg5}-\ref{spg8} exhibit magnitudes of $I_c$ and 
$B$ at a typical bias voltage both in case of equal and unequal hoppings 
for these two generations. From Figs.~\ref{spg5} and \ref{spg7} we see 
that mirror symmetry is preserved and it is obtained for any bias voltage
which we confirm through our numerical calculations. Thus it is assured that 
this feature holds true for any generation. As soon as position of leads 
are changed the symmetry gets destroyed which is reflected from 
Figs.~\ref{spg6} and \ref{spg8}, and thus revealing a new feature that 
could not be detected in case of transport current. With increasing
SPG generation both the loop current and magnetic field will decrease in 
magnitude. 

Up to this we consider two types of hopping integrals 
(viz, $t_x$ and $t_y$) to discuss characteristic properties of circular 
current and associated magnetic field in different triangular plaquettes
of SPG networks. From these results we establish when the mirror symmetry
persists and under which situation it disappears. But, one question 
naturally comes how the mirror symmetry becomes affected if one considers
the effect of two distinct hopping integrals (i.e., $t_{y1}$ and $t_{y2}$)
along two different angular directions associated with two different angles, 
together with the horizontal hopping term $t_x$. To address this question
in Figs.~\ref{nfig1} and \ref{nfig2} we present the results of circular 
currents and associated magnetic fields in different triangular plaquettes
for a $6$-site SPG network considering a particular set of parameter 
values. Two different arrangements of $t_{y1}$ and $t_{y2}$ are taken into 
account, those are schematically shown in Figs.~\ref{nfig1}(a) and 
Figs.~\ref{nfig2}(a), respectively, to have a complete idea about the 
mirror symmetry exhibited by $I_c$ and $B$. Very interestingly, we can see
by comparing the results given in Figs.~\ref{nfig1} and \ref{nfig2} that
the mirror symmetry about an imaginary vertical line (passing through  
lattice sites $2$ and $6$) dividing the SPG into two equal halves exists
only when the geometry itself is mirror symmetric about this imaginary
line. This behavior is clearly reflected in any higher generation which 
we confirm through our extensive numerical calculations. Our
analysis of circular and transport current gives a clear and interesting 
picture of the phenomena occurring in each individual triangles of the 
network.

From the presented results we see that the magnetic field originating at 
the center of each plaquette as a result of bias induced circular current 
exhibits several interesting patterns. For all the three generations 
studied here the magnitude of $B$ is much higher than $7\;$mT in almost all 
the loops except for few triangles, and it is very interesting since such 
an amount of magnetic field is sufficient to cause spin flipping~\cite{Lidar}. 
Therefore, if a spin is situated at the center of the plaquettes, induced 
magnetic field is sufficient enough to cause spin flipping and this 
information might be quiet useful in quantum computation and in spintronics 
applications. We hope by placing magnetic sites in individual plaquettes
and considering the interaction of these local sites with the circular
current induced magnetic fields efficient spin-polarized device
can be designed in the near future.

\section{Conclusion}

We have investigated the phenomena of circulating currents in individual
loops along with overall conductance of a Sierpinski gasket fractal in 
presence of external bias. The effect of self-similar structure
on circular currents and associated magnetic fields have been studied 
considering different generations of the SPG network, and, we have found
that for a particular source-SPG-drain configuration these quantities
reveal definite regularity when the generation gets changed. Most 
importantly we have noted that the model provides mirror symmetry of
current as well as magnetic field with respect to a plane perpendicular 
to the base line (as shown in Fig.~\ref{spg1}), but this symmetry gets 
destroyed for other lead-SPG-lead configurations. Certainly it reveals 
a new feature which could not be detected by analyzing the junction 
current. In our discussion, we have 
considered both the isotropic and anisotropic cases, where the anisotropy
has been introduced through the nearest-neighbor hopping integral. Quite
interestingly we have noted that the inclusion of anisotropy in hopping
integral causes an enhancement of overall junction current, which is 
quite different compared to the conventional systems where anisotropy
or inhomogeneity causes a reduction of net current.

\end{document}